\documentclass[showpacs,preprintnumbers,amsmath,amssymb,aps,onecolumn]{revtex4-1}

\usepackage{graphicx}
\usepackage{bm}
\usepackage{hyperref}
\usepackage{cases}
\usepackage{color}

\begin{document}


\title{Non-collinear interaction of photons with orbital angular momentum}

\author{Thomas Roger$^1$}\email{Corresponding t.w.roger@hw.ac.uk}
\author{Julius J. F. Heitz$^1$}
\author{Ewan M. Wright$^2$}
\author{Daniele Faccio$^1$}
\affiliation{$^1$School of Engineering and Physical Sciences, Heriot-Watt University, EH14 4AS Edinburgh, UK}
\affiliation{$^2$College of Optical Sciences, The University of Arizona, Tucson, Arizona 85721, USA}


\date{\today}

\begin{abstract}
We study the nonlinear interaction between two non-collinear light beams that carry orbital angular momentum (OAM). More specifically, two incident beams interact at an angle in a medium with a second order nonlinearity and thus generate a third, non-collinear beam at the second harmonic frequency that experiences a reduced conversion efficiency in comparison to that expected based on conventional phase-matching theory. This reduction scales with the input beam OAM and, differently from previous spiral bandwidth calculations, is due to a geometric effect whereby the input OAM is projected along the non-collinear interaction direction. The effect is relevant even at small interaction angles and is further complicated at large angles by a non-conservation of the total OAM in the nonlinear interaction.  Experiments are performed under different conditions and are in excellent agreement with the theory. Our results have implications beyond the specific case studied here of second-harmonic generation, in particular for parametric down-conversion of photons or in general for phase-matched non-collinear interactions between beams with different OAM. 
\end{abstract}

\maketitle
\section*{Introduction}
Orbital angular momentum (OAM) is an important degree of freedom in the control of coherent light beams. Of particular interest are nonlinear frequency conversion processes and in general photon-photon interactions in nonlinear media in the presence of OAM. Beams with a spiralling phase front have been used to demonstrate the spatiotemporal entanglement of light in both type-I \cite{Gatti} and type-II \cite{Leach} parametric down conversion, as well as having applications in quantum cryptography \cite{Qian, Boyd2}, providing new imaging applications \cite{Padgett} and improving communication and signal processing protocols \cite{Tamburini}. There has also been a recent surge in interest in using OAM beams for high harmonic generation \cite{Spielman} in the extreme ultraviolet spectral region.\\
The simplest interaction geometries involve a single pump beam and a collinearly generated frequency converted beam. The phase matching properties are not modified by the presence of OAM in the pump beam \cite{Bey,DhoSimPad96} and furthermore OAM is also conserved, e.g. second harmonic photons have an OAM that is twice the OAM of the input pump photons \cite{padgettSHG}. However, many modern applications of nonlinear optics, e.g. coherent control of attosecond pulse generation \cite{Corkum} or the generation of entangled photons, employ a non-collinear interaction.
Entangled photons are created in spontaneous parametric down conversion (PDC) and are, in general, emitted at an angle with respect to the pump.  The ``spiral bandwidth'', i.e.  influence of OAM in the generation and quantum entanglement of noncollinear PDC photons has been considered under the assumption that the phase-matching properties are insensitive to the angle between the OAM beams \cite{TorAleTor03,MiaYaoBar11}. Non-collinear interactions have also been considered using fractional OAM beams and under the assumption that OAM is conserved so as to generate non-trivial OAM states through input beam combination and frequency conversion \cite{Bovino}. However, very large interaction PDC angle geometries have been shown to lead to a violation of the conventionally accepted OAM conservation rule \cite{torner2003}. This violation has a purely geometric origin and can be understood by considering that for large angles, the projection of the output beam onto the non-collinear input pump direction leads to non-integer OAM values, i.e. to the superposition of multiple OAM states \cite{torner2003}.\\
Here we consider the reverse process of PDC, i.e. the case of non-collinear second harmonic generation (SHG) and provide a generalised theory for non-collinear SHG involving OAM beams.  From the perspective of testing the phase-matching theory the SHG geometry considered here offers the benefit that the non-collinear interaction angle can be set at will whereas in PDC, phase-matching determines the emission angles thereby complicating analysis.  We show that when the interaction angle is properly accounted for, the oribital angular momentum of the pump significantly modifies the phase matching relations and leads to a marked reduction in the conversion efficiency. These results also apply to the PDC case and imply a significantly smaller spiral bandwidth than may otherwise be expected, even for angles less than those required for non-conservation of OAM. \\
\section*{Results}
\begin{figure}[h!]
\begin{center}
\includegraphics[width=0.7\textwidth]{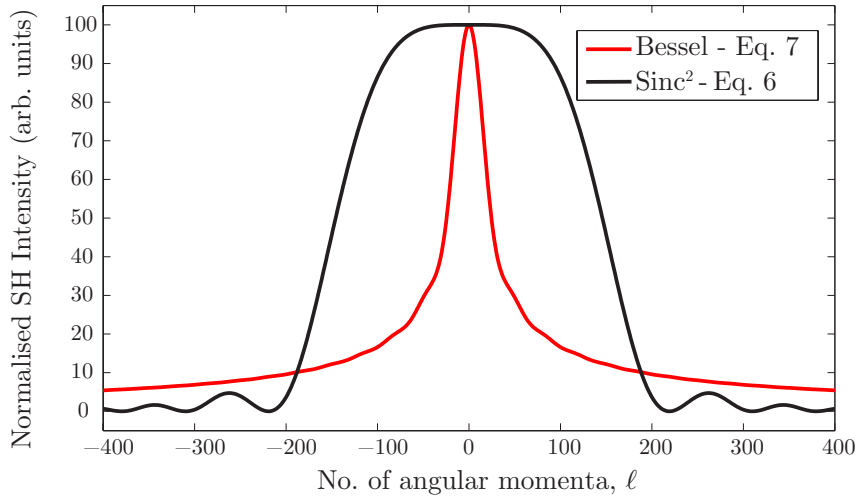}
\caption{Theoretical plot of normalised second harmonic power as a function of the vortex beams winding number $\ell$. The curves are plotted according to the non-collinear interaction of a Gaussian and vortex beam using the phase-matching solutions given by Eq. \ref{eqn:c1s} (black line) and Eq. \ref{eq:c1} (red line). The angle of incidence was chosen to be $\theta = 5^{\circ}$ and all other parameters are the same as in experiments.}
\label{f:theory}
\end{center}
\end{figure}

\textit{Theory}- For purposes of clarity in presentation, we have relegated the detailed description of our phase-matching theory to the Supplementary Information (SI) and the interested reader is referred there.  Instead, in this section our goal is to present the essence of our theory in order to explore its experimental consequences for non-collinear SHG with OAM beams.  To proceed we describe the basic geometry and equations used to treat the non-collinear interaction of OAM beams in a crystal with second order ($\chi^{(2)}$) nonlinearity. In particular, we consider type I, SHG in a BBO crystal.  Two fundamental fields of frequency $\omega$ and wavelength $\lambda_1$, traveling at small angles $\pm\theta$ with respect to the z-axis are labeled $j=0,1$. These are (ordinary) o-waves in the crystal, with the generated second-harmonic produced as an (extaordinary) e-wave, labeled $j=2$, which propagates along the z-axis.  We employ the undepleted pump beam and paraxial approximations. Each fundamental field may be either a Gaussian or a field with a ring shaped intensity profile of radius $R>>\lambda_1$ that carries OAM with winding number $\ell$.  The transverse spatial extent of both fundamental beams is assumed so large that diffraction may be neglected within the medium, that is, their Rayleigh ranges are much larger than the medium length $L$.\\
For an OAM beam with azimuthal variation $e^{i\ell\phi}$ propagating along the z-axis, the corresponding spiraling wave vector may be written as
\begin{eqnarray}\label{K}
\vec{K} &=& K_x\vec{e}_x + K_y\vec{e}_y + K_z\vec{e}_z  \nonumber\\
&=&  {\ell\over R}\cos(\phi)\vec{e}_x + {\ell\over R}\sin(\phi)\vec{e}_y + \sqrt{k_o^2-{\ell^2\over R^2}}\vec{e}_z  ,
\end{eqnarray}
\noindent with $k_o$ the magnitude of the ordinary wave vector for the fundamental field, and $\phi$ the azimuthal angle around the ring.  For use in our non-collinear geometry we rotate this wave vector by a small angle $|\theta|\ll1$ around an axis, that here we take as the y-axis, so that the components of the wave vector become to leading order
\begin{eqnarray}\label{Kprime}
K_x^\prime &=& K_x\cos(\theta)-K_z\sin(\theta) \approx {\ell\over R}\cos(\phi) - k_o\theta , \nonumber \\
K_y^\prime &=& K_y\approx {\ell\over R}\sin(\phi) , \nonumber \\
K_z^\prime &=& K_z\cos(\theta)+K_x\sin(\theta) \approx  k_o - {1\over 2k_o}{\ell^2\over R^2}+{\ell\theta\over R}\cos(\phi). \nonumber \\
&&
\end{eqnarray}
\noindent The rotated wave vector $\vec{K}'$ has an associated plane-wave
\begin{equation}
e^{i\vec{K}'\cdot\vec{r}} \rightarrow e^{iK_z^\prime z} e^{-ik_0\theta x} e^{i\ell\phi}   .
\end{equation}
So as a result of the rotation we have from left to right, a modification to the z-component of the wave vector, a propagation component along the x-axis, and finally the original OAM factor $\exp(i\ell\phi)$.  Consider now the non-collinear interaction between an OAM beam as above and a Gaussian traveling at the opposite angle: we shall refer to this as case (i).
\textit{Case (i)} - Here the second-order polarization $P^{(2)}$ that will drive the SHG will be proportional to the product of the plane-waves as above for the OAM beam, with $\ell$ and $\theta$, and the Gaussian, with $\ell=0$ and $-\theta$, giving
\begin{eqnarray}\label{PWfac}
P^{(2)}&\propto & e^{2ik_o z+ i\ell\phi - {iz\over 2k_o}{\ell^2\over R^2}+{iz\ell\theta\over R}\cos(\phi)} \nonumber \\
&=& e^{2ik_o z+ i\ell\phi - {iz\over 2k_o}{\ell^2\over R^2}}
\sum_{n=-\infty}^{\infty}i^n e^{in\phi}J_n\left( {z\ell\theta\over R} \right)   ,
\end{eqnarray}
where the Jacobi-Anger expansion of the term $e^{{iz\ell\theta\over R}\cos(\phi)} $ has been employed in the bottom line.  In contrast to previous work, here we explore the Bessel function terms appearing in Eq. (\ref{PWfac}), which capture the dependence of the nonlinear interaction on the angle between the OAM beams.  We note that although the input fields have OAM of $\ell$ and zero (the Gaussian) in our example, the nonlinear polarization for the SHG in Eq. (\ref{PWfac}) contains many OAM components of the form $e^{i(\ell+n)\phi}$ weighted by the Bessel functions.  This reflects the fact that OAM need not be conserved during non-collinear interactions between beams carrying OAM for large enough angles.  Here we confine our attention to the case $n=0$ corresponding to SHG with the same winding number $\ell$ as the input fundamental, giving
\begin{equation}\label{PWfac0}
P^{(2)}\propto e^{2ik_o z+ i\ell\phi - {iz\over 2k_o}{\ell^2\over R^2}}J_n\left( {z\ell\theta\over R} \right)   ,
\end{equation}
which is justified if the argument $|z\ell\theta/R|<1$ remains small since $J_n(s)\rightarrow 0$ for $n\ne 0$ and small $|s|$.  In conventional small angle theories of non-collinear phase matching only the exponential term in Eq. (\ref{PWfac}) has been considered, and this leads to a variation of the SHG power with crystal length $L$ of the well known sinc$^2$ form
\begin{equation}\label{eqn:c1s}
P_2(\ell) \propto sinc^2\left ({\ell^2\over 16\pi}{\lambda_1 L\over n_oR^2} \right )  ,
\end{equation}
with $n_o$ the ordinary refractive-index.  We introduce the critical winding number $\ell_C=4\pi R\sqrt{n_o/\lambda_1L}$ as that for which the SHG power falls to zero according to Eq. (\ref{eqn:c1s}).  This result is closely related to and follows the spiral bandwidth relations derived in previous works \cite{MiaYaoBar11, McClaren, Alarcon}.  In contrast, in the opposite extreme that we retain only the Bessel term in Eq. (\ref{PWfac0}) then the SHG output power assumes the form
 \begin{equation}
P_2(\ell) \propto\left |\int_0^L dz J_0\left( {z\ell\theta\over R} \right )\right |^2 ,
\label{eq:c1}
\end{equation}
and we introduce the critical winding number $\ell_c\approx |3R/\theta L|$ as that at which the SHG power drops to zero according to Eq. (\ref{eq:c1}) (see SI for details).   Then the ratio of the two critical winding numbers 
\begin{equation}
r={\ell_C\over \ell_c} \simeq 4|\theta|\sqrt{{Ln_o\over \lambda_1}}  .
\label{eq:c1b}
\end{equation}
tells us whether the conventional exponential phase-matching term is dominant ($r<<1$) or the new Bessel term ($r>>1$).  To illustrate the differences that can arise from conventional exponential phase-matching term versus the new Bessel term, in Fig.~\ref{f:theory} we show a comparison between these two regimes for a non-collinear angle of $\theta = 5^{\circ}$.  This example shows that the Bessel term dependence on the OAM via $\ell$ can greatly reduce the SHG power, and hence conversion efficiency, thereby also greatly reducing the available spiral bandwidth.\\
We consider three cases of phase-matching with vortex beams: (i) one OAM beam with winding number $\ell$ and a Gaussian ($\ell=0$) beam, that we discussed above, (ii) two OAM beams with opposite winding numbers $\ell$ and $-\ell$ respectively, and (iii) two OAM beams, both with the same winding number $\ell$.
\begin{figure}[t!]
\begin{center}
\includegraphics[width=0.7\textwidth]{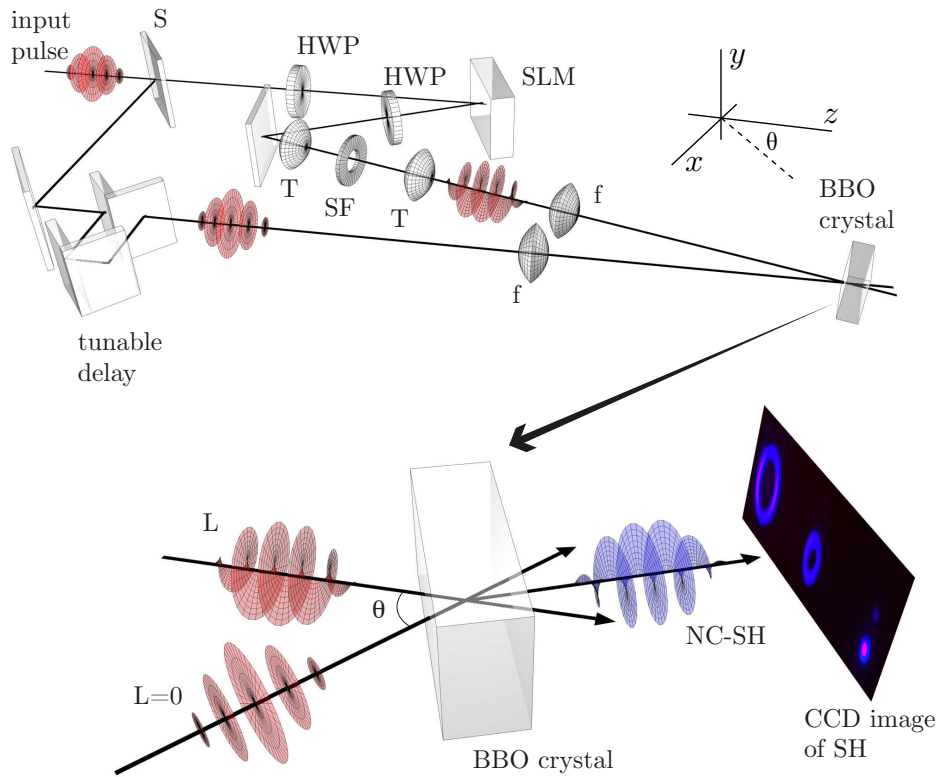}
\caption{Experimental layout for \textit{case (i)}, described in detail in the methods section. Input pulses are split by a beamsplitter (S), the reflected component, with Gaussian profile (L=0), passes a tuneable delay stage while the second traverses a fixed optical path in which the pulse's phase and amplitude are corrected (by a series of half-wave plates (HWP) and a spatial light modulator (SLM)) to impart OAM onto the beam (L). A spatial filter (an aperture (SF) in the fourier plane of a 1:1 telescope (T)) removes higher order OAM modes and the beams (L=0 and L) are loosely focussed (F) and overlapped non-collinearly onto a BBO crystal. The non-collinear output is measured with a photodetector.}
\label{f:layout}
\end{center}
\end{figure}

\noindent
\textit{Case (ii)} - In this case two fundamental beams of opposite OAM combine to make a SH field with zero winding number (to leading order).  This case is therefore the reverse process of degenerate parametric down conversion so lessons learned here may also apply to PDC \cite{TorAleTor03,MiaYaoBar11}.  In particular, we find that this case involves not only the conventional phase-matching described by the exponential term but also has the Bessel term, as seen in \emph{case (i)}. To the best of our knowledge the Bessel term has not appeared in previous treatments of PDC, perhaps justified by the fact that small angles are employed: In the limit $\theta \rightarrow 0$ the Bessel term tends to unity and only the conventional exponential term remains. In that limit, we assume the SHG is phase-matched for $\ell=0$, which requires $\Delta k=(2k_0-k_e)=0$,  and the SHG power varies as Eq. (\ref{eqn:c1s}), with an additional factor 4 in the argument of the \emph{sinc} function (see SI).
For large $r$, we may retain the Bessel term alone which again cannot be represented in a closed form
\begin{equation}
P_2(\ell) \propto \left |\int_0^L dz J_0\left( {2z\ell\theta\over R} \right )\right |^2.
\label{eq:c2}
\end{equation}
This is of the same form as in \textit{case (i)} but now with a factor 2 larger argument of the Bessel function.\\
\textit{Case (iii)}- For a fixed medium length the SH power will vary with the winding number $\ell$ as
\begin{equation}
P_2(\ell)\propto sinc^2\left ({\Delta k(\ell)L\over 2} \right ) .
\end{equation}
If we assume that the SHG is phase-matched for $\ell=0$ which again requires $\Delta k=(2k_0-k_e)=0$, then using $k_e=2k_o$ we find $\Delta k(\ell)=0$.  This is in keeping with the known result for single beam SHG that if the integration is phase-matched for a Gaussian beam it is also phase-matched for vortex beams of varying winding number \cite{DhoSimPad96} (or indeed any superposition of Laguerre-Gauss beams). Thus the SHG power is independent of the winding number $\ell$.
\begin{figure*}[ht]
\begin{center}
\includegraphics[width=\textwidth]{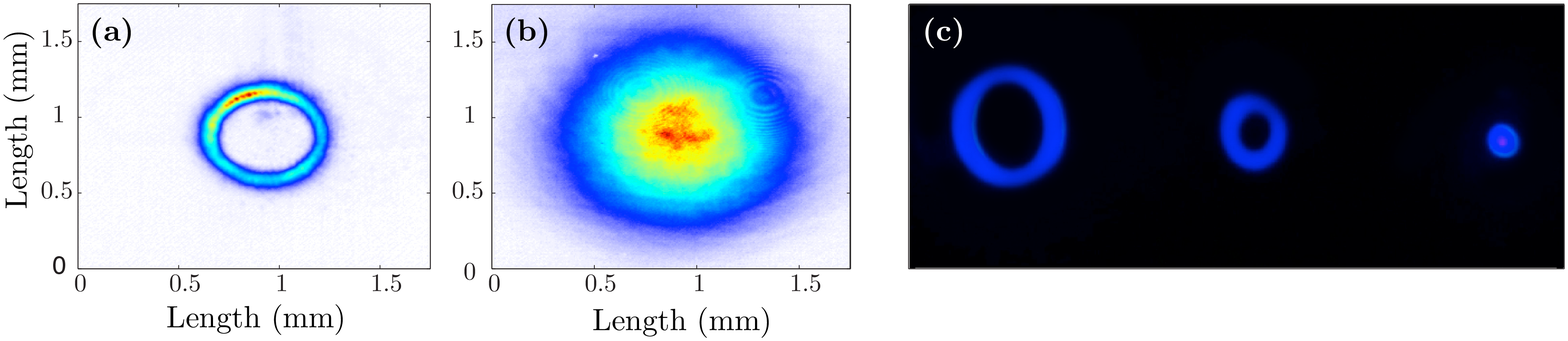}
\caption{Spatial intensity profiles of the (a) spiral ($\ell = 20$) and (b) Gaussian beams at the overlap position (within the BBO crystal) measured with a beam profiler. The spiral and Gaussian beams have diameters of $\sim 0.4$ mm and $\sim 1.2$ mm respectively. The size of the spiral beam is kept constant as the OAM number $\ell$ is varied. (c) a photograph of the second harmonic (SH) generated beams projected onto a surface shortly after the BBO crystal position, from left to right the spiral SH, the non-collinear SH and Gaussian SH.}
\label{f:profiles}
\end{center}
\end{figure*}

\begin{figure*}[ht]
\begin{center}
\includegraphics[width=\textwidth]{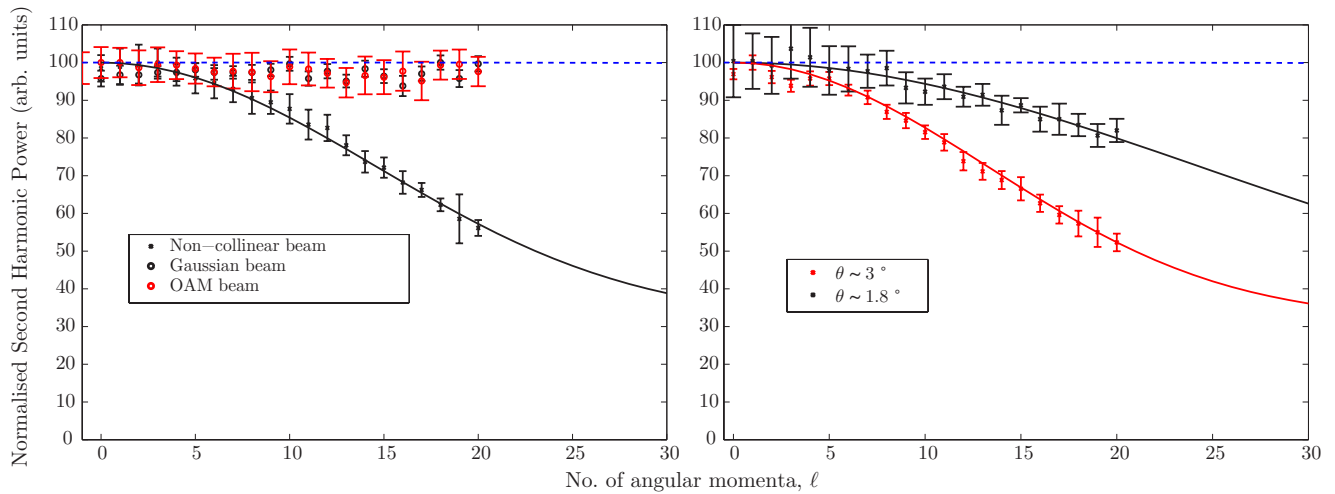}
\caption{Non-collinear second harmonic power as a function of the winding number $\ell$ of the OAM beam. (a) A Gaussian and vortex beam are overlapped in a non-collinear geometry onto a BBO crystal (\emph{Case (i)}). The angle of incidence between the two non-collinear beams was 6$^{\circ}$. (b) Two vortex beams of opposing winding numbers $\ell$ and $-\ell$ respectively, are overlapped onto a BBO crystal (\emph{Case (ii)}). The non-collinear second harmonic beam is detected. The data are fit using Eq.~(\ref{eq:c1}) in (a) and Eq.~(\ref{eq:c2}) in (b) using only the experimental conditions. The dashed blue line shows the dependence assuming collinear geometry according to Eq.~(\ref{eqn:c1s}). Note that the collinearly detected beams do not vary with winding number (only shown in (a)).}
\label{f:expt}
\end{center}
\end{figure*}

\textit{Experiment}- Experiments are performed in order to measure the predicted change in the phase matching conditions when using spiral beams. Specifically we look at cases \emph{(i)} and \emph{(ii)} described above (\emph{case (iii)} shows no dependency on $\ell$). A standard non-collinear second harmonic generation (SHG) geometry is implemented in which two beams are overlapped onto a SHG crystal. The experimental scheme for \emph{case (i)} is shown in Fig. \ref{f:layout} (see methods). The output of an amplified Ti:Sapphire pulsed laser, producing $\sim$90 fs pulses, centred at $\lambda_0 = 785$ nm with a repetition rate of 100 Hz, is split into two equal components by a 50:50 beamsplitter. A spiral phase is applied to one of the beams via a spatial light modulator (SLM) \cite{Padgett}. The beams are loosely focussed to spatially overlap the beams at the BBO crystal at an angle $\theta = 6 \pm 1^{\circ}$.
The input and output beam profiles are shown in Fig. \ref{f:profiles}. We note that the size of the beams does not appreciably change either over the interaction length ($L_{int} \sim$ 500 $\mu$m) or for varying OAM within the values $\ell$ used in our experiments. We calculate that the interaction length is limited by group velocity mismatch (GVM) and accordingly we use a BBO crystal of the same length.
The beams are detected after the BBO crystal by a photodetector with a blue filter used to reject the fundamental and measure only the second harmonic light. An aperture isolates the non-collinear beam and the power is measured as the winding number is varied between $\ell = 0\ldots20$, shown in Fig. \ref{f:expt} (a). The most obvious feature is that whilst the collinear second harmonic power of the two  beams do not change with $\ell$ (in agreement with previous work), there is a marked $\sim$ 40 $\%$ decrease in the SH power of the non-collinear beam as the OAM is tuned. The experimental data is also fitted very precisely using Eq.~(\ref{eq:c1}) with no free parameters.\\

We extend this study using a similar experimental setup in order to study \emph{case (ii)} presented above. The filtered output of the SLM is now split into two equal components by moving the 50:50 beamsplitter, S, after the telescope. Two oppositely spiralling OAM beams are obtained by placing and even number and odd number of mirror reflections on the two separate arms. The two beams are then overlapped non-collinearly onto the BBO sample at a non-collinear incidence angle, $\theta$. Two different experiments where performed in which we investigated the non-collinear conversion efficiency for two different interaction angles, $\theta = 3 \pm 0.5^{\circ}$ and $\theta = 1.8 \pm 0.2^{\circ}$. 
\noindent We measure a $\sim 50 \%$ decrease in SH power for an incidence angle of $\theta = 3^{\circ}$ and $\sim 15 \%$ decrease for $\theta = 1.8^{\circ}$. The normalised SHG power is plotted in Fig. \ref{f:expt} (b) as a function of the OAM winding number $\ell$. The fall-off in the generated second harmonic beam follows Eq. (\ref{eq:c2}) very precisely, for both interactions angles. It is worth noting that the value of the parameter $r = \ell_C / \ell_c$ is much greater than 1 ($r_{exp} \geq 4$) for all the experimental conditions used and thus we do indeed expect to see the Bessel dependence as per Eq.~(\ref{eq:c2}). \\

\section*{Discussion}

We have demonstrated experimentally that for the case of small angles that non-collinear SHG with vortex beams can be accurately described by a generalized phase-matching theory that yields Eqs.~(\ref{eq:c1}) and (\ref{eq:c2}) as limiting cases. However as alluded to earlier, the full description of non-collinear phase matching is described by a more complex set of equations (see SI). In particular, the second-order nonlinear polarization is proportional to the product of the two fundamental fields, and may be written for non-collinear interaction between two fundamental fields $E_0(\vec{r})$ and $E_1(\vec{r})$ as (see SI)
\begin{widetext}
\begin{numcases}{E_0(\vec{r})E_1(\vec{r}) =}
A_G(\rho)A_\ell(\rho)\sum_{n=-\infty}^{\infty}i^n\exp\left [ i(\ell+n)\phi - {iz\over 2k_o}{\ell^2\over R^2} \right ]J_n\left( {z\ell\theta\over R} \right)  & for \text{case (i)}, \label{eqn:f1}
\\
A_\ell^2(\rho)\sum_{n=-\infty}^{\infty}i^n\exp\left [ in\phi - {iz\over k_o}{\ell^2\over R^2} \right ]J_n\left( {2z\ell\theta\over R} \right)  & for \text{case (ii)}. \label{eqn:f2}
\end{numcases}
\end{widetext}
We find that in \emph{case (i)} (Eq.~\ref{eqn:f1}), where $A_G(\rho)$ and $A_{\ell}(\rho)$ are the electric field amplitudes of the Gaussian and vortex beam respectively, that there is an additional phase imparted onto the non-collinear field when $n>0$. In the previous description we assumed that for non-collinear angles between zero and a critical angle $\theta_c$, we use only the zero order terms (i.e. $n = 0$). This critical angle may be determined by considering the situation for which the SH power goes to zero, corresponding to some critical winding number $\ell_{max}$. In our measurements we have a maximum value of $\ell = \ell_{max} = 20$ and we can then find the maximum value of $\theta = \theta_{max}$ for each case: these are the limiting interaction angles above which, we should expect a relevant role of the $n>0$ terms. Thus, $\theta_{max} = 3L/\ell_{max}R \sim 7^{\circ}$ and $\theta_{max}= 3L/2\ell_{max}R \sim 3.5^{\circ}$ for cases \textit{(i)} and \textit{(ii)}, respectively. If we were to include the full description of the non-collinear phase matching for higher angles we would expect to see fractional OAM winding numbers appearing in the non-collinear beam. This is of critical importance when considering the generation of high order winding numbers or indeed PDC at large angles in entanglement experiments as it suggests that OAM modes generated in these cases will not conserve the input OAM \cite{torner2003}.
Finally, we note that an expression for the SHG power of the same form shown here will arise in the treatment of walk-off effects and beam tilt due to the SLM but with $\theta$ in the expression for $\ell_c = |3R/\theta{L}|$ replaced by the walk-off angle or tilt angle. \\

\section*{Methods}
The experimental scheme to measure the effect of OAM on phase matching in BBO is presented in Fig. \ref{f:layout}. The figure and the following paragraph describe the experiment performed to explore \textit{case(i)} outlined in the \emph{theory} section. Ultrashort pulses of $\Delta \tau \sim 90$ fs are input at 100 Hz, the pulses are split by a 50:50 beamsplitter into two equal components with flat phase fronts. The transmitted beam is polarised with respect to a spatial light modulator (SLM) by rotation of a half-wave plate (HWP). The SLM imparts a phase and amplitude correction to the beam in order to produce spiral pulses with $\ell = -20\ldots20$ with equal diameter. In order to select the desired $\ell$ value a diffraction grating is superimposed onto the SLM to separate spatially the various $\ell$ modes, which are selected in the Fourier plane of a 4f telescope (T) using and spatial filter (SF). The reflected beam of the beamsplitter is delayed with respect to the other beam by a 1-dimensional translation stage. Both beams (L and L=0) are loosely focussed (f) toward the BBO crystal in a non-collinear geometry at an angle of $\theta = 6^{\circ}$. The output beams, including the generated resultant non-collinear second harmonic (NC-SH) beam, are measured following the BBO with a photodetector. For \emph{case (ii)} the beamsplitter (S) is moved after the spatial filter and an odd or even number of mirror reflections control the sign of the OAM number ($\ell, -\ell$) on the beams.

\begin{acknowledgments}
D.F. acknowledges  financial support from the Engineering and Physical Sciences Research Council EPSRC, Grant EP/J00443X/1 and from the European Research Council under the European Union's Seventh Framework Programme (FP/2007-2013) / ERC Grant Agreement n. 306559. E.W. acknowledges financial support from the Scottish Universities Physics Alliance, SUPA.
\end{acknowledgments}

\section*{Author Contributions}
D.F. developed the ideas and lead the project. E.W. developed the theoretical model. T.R. and J. H. performed the experimental measurements. All authors contributed to the manuscript.

\section*{Competing Financial Interests}
The authors declare no competing financial interests.


%

\end{document}